# SALIENCE: An Unsupervised User Adaptation Model for Multiple Wearable Sensors Based Human Activity Recognition

Ling Chen, Yi Zhang, Shenghuan Miao, Sirou Zhu, Rong Hu, Liangying Peng, and Mingqi Lv

**Abstract**—Unsupervised user adaptation aligns the feature distributions of the data from training users and the new user, so a well-trained wearable human activity recognition (WHAR) model can be well adapted to the new user. With the development of wearable sensors, multiple wearable sensors based WHAR is gaining more and more attention. In order to address the challenge that the transferabilities of different sensors are different, we propose SALIENCE (unsupervised u̲s̲er a̲daptation mode̲l̲ for mult̲i̲ple w̲e̲arable se̲n̲sors based human a̲c̲tivity r̲e̲cognition) model. It aligns the data of each sensor separately to achieve local alignment, while uniformly aligning the data of all sensors to ensure global alignment. In addition, an attention mechanism is proposed to focus the activity classifier of SALIENCE on the sensors with strong feature discrimination and well distribution alignment. Experiments are conducted on two public WHAR datasets, and the experimental results show that our model can yield a competitive performance.

**Index Terms**—Human activity recognition, multiple wearable sensors, unsupervised domain adaptation

—————————— ◆ ——————————

## 1 INTRODUCTION

WEARABLE human activity recognition (WHAR) infers users' activities using the wearable sensors attached to different human body parts, which is one of the key tasks in the field of ubiquitous and mobile computing. Currently, a variety of applications are supported by WHAR, e.g., health support [15] and worker assistance [28].

User generalization has always been a key challenge for WHAR. Due to the unique activity patterns of each individual user [6, 17], there is a significant distribution shift between the data collected from different users. We give an example in Section 4.5 by visualizing the feature distributions of the new user and training users. Applying the models trained on training users to a new user typically results in significant performance degradation. To meet this challenge, personalized WHAR is proposed, which builds a tailored model for each user based on their unique activity patterns [7, 16, 19, 29, 42]. Collecting a large amount and variety of labelled data to train a personalized model for each new user is usually costly. Some researches use a small amount of labelled data from the new user to adapt a general model to a personalized model [29, 40, 42, 55]. Since the collection of labelled data typically requires user participation, this limits the use of these methods. Considering that unlabelled data are easier to obtain, some researches attempt to adapt a general model to a personalized model using only the unlabelled data of the new user [7, 16, 19]. They align the feature distributions of the data from training users and the new user using unsupervised domain adaptation (UDA), which can be achieved by various approaches, e.g., minimizing the KL divergence [19], optimizing the Maximum Mean Discrepancy (MMD) [16], and using a domain-adversarial neural network [7], so a well-trained model can be well adapted to the new user. In this paper, we focus on the unsupervised user adaptation problem, i.e., using labelled data from training users and unlabelled data from the new user to train a WHAR model for the new user.

Many early researches [5, 9, 11, 21, 22, 51] use a single accelerometer to recognize simple human activities, e.g., standing, sitting, and lying. With the development of wearable sensors, the acquired sensing data are becoming more and more abundant, covering not only different body parts, but also different modalities, which can support recognizing more diverse and complex activities and bring more accurate recognition results. However, the use of multiple wearable sensors poses a challenge for unsupervised user adaptation: The transferabilities of different sensors are different. Existing methods usually do not take this difference into consideration, and simply concatenate the data from different sensors into a whole to perform unsupervised user adaptation, which are effective in ensuring global alignment, but cannot achieve

————————————————

- L. Chen, Y. Zhang, S. Miao, S. Zhu, R. Hu, and L. Peng are with the College of Computer Science and Technology, Zhejiang University, Hangzhou 310027, China, and also with Alibaba-Zhejiang University Joint Research Institute of Frontier Technologies, Hangzhou 310027, China.
- M. Lv is with the College of Computer Science and Technology, Zhejiang University of Technology, Hangzhou 310023, China.
- E-mail: {lingchen, zhangyi1995, shmiao, 3150105210, ronghu, lyoare}@zju.edu.cn, mingqilv@zjut.edu.cn.

local alignment.

To address the above challenge, we propose SALIENCE (unsupervised user adaptation model for multiple wearable sensors based human activity recognition) model. To achieve local alignment, it performs alignment at the sensor level based on the unique data distribution of each sensor. Considering the different alignment degrees of sensors, it differentiates the importance of different sensors. The main contributions of this paper are summarized as follows:

1) Propose an alignment scheme for multiple sensor data, which aligns the data of each sensor separately to achieve local alignment, and uniformly aligns the data of all sensors to ensure global alignment.

2) Propose an attention mechanism to focus the model on the sensors with strong feature discrimination and well distribution alignment, which further improves the model performance and generalization.

3) Evaluate SALIENCE on two public WHAR datasets, and the results show that our model has competitive performance in multiple wearable sensors based WHAR.

The remainder of this paper is organized as follows: Section 2 reviews the related work; Section 3 describes the proposed model in detail; The experiments are shown in Section 4; Finally, Section 5 concludes this paper.

## 2 RELATED WORK

### 2.1 Wearable Human Activity Recognition

WHAR infers users' activities using the wearable sensors attached to different human body parts. Commonly used wearable sensors include accelerometer, gyroscope, magnetometer, and heart rate meter. According to the number of sensors used, WHAR can be divided into three categories, i.e., using single wearable sensor [5, 9, 11, 21, 22, 27, 51], using multi-modal wearable sensors on single body position [12, 13, 23], and using multi-modal wearable sensors on multiple body positions [1, 3, 8, 33, 35, 48]. Due to the limited information obtained by a single sensor, single wearable sensor based WHAR is typically used to recognize simple human activities, e.g., standing, sitting, and lying. For example, Kwapisz et al. [22] exploited a phone-based accelerometer to recognize walking, jogging, ascending stairs, descending stairs, sitting, and standing. Chen et al. [5] proposed a convolutional neural network (CNN) architecture for WHAR using a single accelerometer. With the development of wearable sensors, multi-modal wearable sensors based WHAR has attracted more and more attention. Multi-modal wearable sensors can obtain rich information, covering not only different body parts, but also different modalities, which supports recognizing more diverse and complex human activities and brings more accurate recognition results. For example, Peng et al. [35] combined physiological sensors and accelerometers attached to different human body parts to recognize complex human activities. Kern et al. [18] investigated the influence of the number of sensors on the performance of WHAR, and the results suggest that combining multi-modal wearable sensors is essential for recognizing complex human activities.

Most WHAR methods mix all labelled activity data from training users, train a model using supervised learning methods, and finally apply the trained model directly to new users [1, 13, 23, 31, 35, 37]. Due to the unique activity patterns of each individual user, these methods face the problem of user generalization. Personalized WHAR is proposed to address this problem, which builds a tailored model for each user based on their unique activity patterns. Since it is difficult to ask each user to provide rich and diverse labelled data, personalized WHAR typically uses some data from the new user for user adaptation, which can be divided into two categories: supervised user adaptation and unsupervised user adaptation. Supervised user adaptation uses a small amount of labelled data from the new user to adapt a general model to a personalized model [29, 40, 42, 55]. For example, Rokni et al. [42] proposed a supervised user adaptation method based on deep transfer learning, which reuses the lower layers of a well-trained WHAR network and fine-tunes the upper layers with only a small amount of labelled data from the new user. Matsui et al. [29] developed a CNN based supervised user adaptation method for WHAR, which inserts a special layer with a small number of free parameters between each two convolutional layers and optimizes these free parameters using a small amount of labelled data from the new user. Since the collection of labelled data requires user participation, this limits the use of supervised user adaptation. Unsupervised user adaptation adapts a general model to a personalized model using only unlabelled data from the new user, which is realized by UDA [7, 16, 19]. Considering that unlabelled data are easier to obtain, it is more practical than supervised user adaptation.

### 2.2 Unsupervised Domain Adaptation

Due to the phenomenon known as domain shift [38], models trained on the source domain do not generalize well to the target domain. Domain adaptation is a machine learning technique that adapts models between domains. In the case of UDA, no labelled data are available in the target domain. Most existing methods perform UDA by aligning the distributions of the source and target domain in a common feature space. According to the strategies used, these methods can be roughly divided into three categories. The first category optimizes the measures of distributional discrepancy [25, 43, 46]. MMD is the most commonly used measure [25, 43], which calculates the distance between the means of two domains in some reproducing kernel Hilbert spaces. For example, Rozantsev et al. [43] used MMD to measure the distributional discrepancy. Sun and Saenko [46] used the distance between the second-order statistics (covariances) of the source and target features as the measure of distributional discrepancy. The second category introduces a domain discriminator that distinguishes two domains, and adversarially optimizes the domain discriminator and feature extractor to encourage the feature extractor to learn domain-invariant features [10, 34, 49]. The third category transforms the distribution of the source domain to resemble the distribution of the target domain with generative models [2, 14, 47]. Considering the excellent data generation capabilities, generative adversarial network (GAN) is often used. For example, Bousmalis et al. [2] used GAN to learn a mapping that transforms the data in the source domain into data that appear to be sampled from the target domain while retaining their original information.

UDA is naturally used for unsupervised user adaptation in WHAR, which uses the rich and diverse labelled data from training users to learn an accurate WHAR model for the new user whose labelled data are unavailable [7, 16, 19]. For example, Khan et al. [19] used KL divergence to measure the differences between the feature distributions of labelled data from training users and unlabelled data from the new user, and aligned the feature distributions by minimizing the KL divergence. Hosseini et al. [16] performed unsupervised user adaptation by optimizing MMD. Ding et al. [7] used a domain-adversarial neural network model for unsupervised user adaptation in WHAR, which consists of a feature extractor, an activity classifier, and a domain discriminator. The domain discriminator aims to distinguish whether a sample is from training users or the new user, while the feature extractor tries to fool the domain discriminator. By optimizing the domain discriminator and the feature extractor alternately, the differences between the feature distributions of the data from training users and the new user are continuously reduced. The feature extractor eventually learns features that are difficult to distinguish even with a well-trained domain discriminator, so the well-trained activity classifier can be well adapted to the new user. Sanabria et al. [45] used bi-directional GAN for heterogeneous feature transfer and contrastive learning to minimise the intra-class discrepancy and maximise the inter-class margin. Liu et al. [24] proposed adversarial spectral kernel matching method that can precisely characterize non-stationary and non-monotonic statistics in time series distributions. These methods directly apply general UDA methods to WHAR, ignoring the challenge posed by multiple sensor data: The transferabilities of different sensors are different. Different from the existing UDA methods, the UDA method proposed in this paper can achieve both local alignment and global alignment. The former is achieved by aligning the data of each sensor separately and the latter is achieved by uniformly aligning the data of all sensors.

## 3 METHODOLOGY

### 3.1 Problem Definitions

For multiple wearable sensors based WHAR, it is assumed that there are $K$ wearable sensors attached to different human body parts to collect activity data. Raw streaming data are pre-processed and segmented into equal-sized time windows as input data for WHAR. A time window $x$ can be formulated as:

$$x = [r_1, r_2, \cdots, r_k, \cdots, r_K], k \in \{1, 2, \cdots, K\} \quad (1)$$

where $r_k \in R^{l \times c_k}$ represents the segmented record matrix of sensor $k$, $l$ is the number of records within a time window, and $c_k$ denotes the channel number of sensor $k$. Let $\mathcal{X}$ represent the input space, $\mathcal{Y}$ represent the set of activity labels, $x^{TU} \sim P$ denote the data from training users and $x^{NU} \sim Q$ denote the data from the new user, $P$ and $Q$ are different distributions. The labelled dataset of training users $D^{TU}$ and the unlabelled dataset of the new

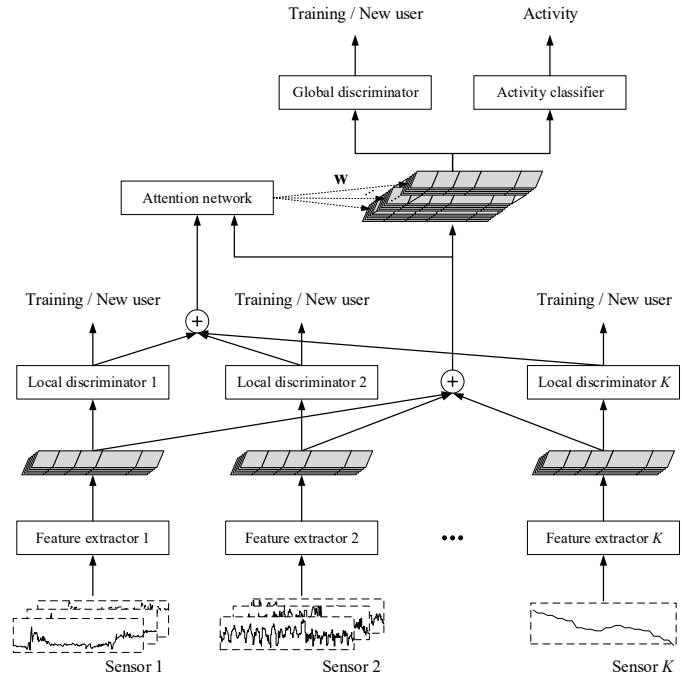

Fig. 1. The framework of SALIENCE. Solid rectangles represent neural networks, stacked parallelograms represent deep features, "⊕" represents concatenation operation, and arrows indicate data flow.

user $D^{NU}$ can be formulated as:

$$D^{TU} = \{(x^{TU}, y)\}, x^{TU} \in \mathcal{X}, y \in \mathcal{Y} \quad (2)$$
$$D^{NU} = \{x^{NU}\}, x^{NU} \in \mathcal{X} \quad (3)$$

Given the labelled dataset of training users $D^{TU}$ and the unlabelled dataset of the new user $D^{NU}$, the goal of SALIENCE is as follows: 1) Learn a composite mapping $f \circ g: \mathcal{X} \to \mathcal{Y}$ using $D^{TU}$. Here, function $g(\cdot)$ is used to map the input space to the feature space, and function $f(\cdot)$ is used to map the feature space to the label space. For any unseen sample from training users $(x^{TU}, y)$, $f(g(x^{TU}))$ and $y$ should be as similar as possible; 2) Align the feature distributions of $D^{TU}$ and $D^{NU}$ to make $g(x^{TU})$ and $g(x^{NU})$ obey the same distribution as much as possible. In this way, the learned composite mapping $f \circ g$ can be well applied to the new user to perform WHAR.

### 3.2 Framework

Fig. 1 presents the framework of SALIENCE, which consists of feature extractors, local discriminators, global discriminator, attention network, and activity classifier. Each feature extractor is responsible for processing the data from a separate wearable sensor to extract deep features. Each local discriminator corresponds to a feature extractor, which predicts whether an input sample is from training users or the new user based on the output of the corresponding feature extractor. Attention network assigns weights to sensors to differentiate their importance. Global discriminator predicts whether an input sample is from training users or the new user based on the weighted outputs of all feature extractors. Activity classifier performs WHAR based on the weighted outputs of all feature extractors. In SALIENCE, feature extractors and discriminators are optimized using an adversarial strategy. Discriminators aim to distinguish whether a sample is from training users or the new user, while feature extractors try to fool discriminators. By optimizing these two parts alternately, feature extractors will continuously reduce the differences between the feature distributions of data from training users and the new user. The details of feature extractors, local and global discriminators, attention network, and activity classifier are introduced in Sections 3.3-3.6, and the training process is presented in Section 3.7.

### 3.3 Feature Extractors

Feature extractors are responsible for extracting deep features from sensor data. Considering that multiple wearable sensors attached to different human body parts have large differences in placement, orientation, data modalities, etc., showing different data distributions, the data of each sensor are processed using an independent feature extractor. SALIENCE uses CNNs as feature extractors. We first briefly introduce CNN, and then explain how to use it as a feature extractor.

CNN is a class of feedforward artificial neural network. A CNN architecture is formed by a stack of distinct layers that transform the input through a differentiable function and pass the output to the next layer. Convolutional layer, pooling layer, and activation layer are commonly used. The core of convolutional layer is a set of learnable convolution kernels, which convolve across some dimensions of the input and generate feature maps. The network learns the convolution kernels that are activated when certain types of features in the input are detected. By stacking multiple convolutional layers, abstract features will be extracted layer by layer. Pooling layer is a form of non-linear down-sampling, which can reduce the spatial size of feature maps and control overfitting. There are several non-linear functions to implement pooling, of which max pooling and average pooling are the most commonly used. Activation layer applies non-linear activation function to its input, which increases the non-linear properties of the overall network. Rectified linear unit (ReLU) is one of the most commonly used activation functions, which can be given formally by:

$$\sigma(x) = \max(x, 0) \quad (4)$$

In SALIENCE, feature extractors corresponding to different sensors have different parameters but the same architecture, so we focus on feature extractor $k$ as an example to illustrate the detailed architecture. As shown in Fig. 2, a feature extractor contains three convolutional layers. The first convolutional layer uses 16 2D convolution kernels of size 3 × 5 to extract the cross-channel features of each sensor (a typical 3-axis accelerometer has x, y, and z axis). Then, the other convolutional layers use 16 1D convolution kernels of size 1 × 5 to extract the temporal features of each sensor.

Although more convolutional layers can theoretically extract more discriminative features, it also increases the network complexity and the difficulty of training. Referring to [4], three convolutional layers are used to extract deep features, which can achieve a balance. Convolution kernels are used for processing sensor data in previous studies [31, 36, 53], which scan input along the temporal dimension to discover the useful features. The output of each convolutional layer is activated by ReLU and input to the next layer.

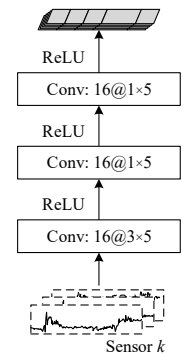

Fig. 2. The architecture of feature extractor $k$. "Conv" refers to convolutional layer. The numbers before and after "@" denote the number and size of convolution kernels, respectively. "ReLU" refers to activation function.

### 3.4 Local and Global Discriminators

Discriminators predict whether an input sample is from training users or the new user based on the outputs of feature extractors, where each local discriminator corresponds to an independent feature extractor and global discriminator corresponds to them all. Temporal dependency discovery is the key to time series data processing. Discriminators exploit bidirectional long short-term memory (bi-LSTM) to mine temporal dependencies and use them for prediction. We first briefly introduce bi-LSTM, and then show the detailed architecture of local and global discriminators.

Recurrent neural network (RNN) is a classic artificial neural network whose previous state is part of its current input. This architecture allows the information to be passed from the current step to the next step, so that the neural network has "memory". Because of its internal memory, RNN is particularly suitable for processing time series data. Bi-LSTM is a specific RNN architecture, which contains two processing units, a forward one and a backward one, to capture temporal dependencies from two directions simultaneously. Each processing unit has a memory cell that is responsible for remembering previous states, and there are three "gates" to control the information flow of memory cell, namely forget gate, input gate, and output gate. These gates output values between 0 and 1 to partially allow or deny information to flow into or out of the memory cell. Forget gate decides the extent to which a value remains in memory cell. Input gate decides the extent to which a value flows into memory cell from input, and output gate decides the extent to which a value in memory cell is output. These well-designed internal structures make it easier for bi-LSTM to learn long-term temporal dependencies.

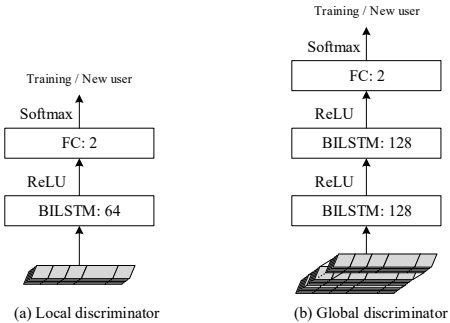

Fig. 3. The architectures of discriminators. "BILSTM" refers to a bi-LSTM layer, where the number represents the size of its state vector. "FC" refers to a fully connected layer, where the number represents the number of its neurons. "Softmax" refers to softmax function.

In SALIENCE, local discriminators corresponding to different feature extractors have different parameters but the same architecture, so we focus on one local discriminator as an example to illustrate the detailed architecture. As shown in Fig. 3, a local discriminator contains one bi-LSTM layer and one fully connected layer. The bi-LSTM layer mines the temporal dependencies in input features, and its output passes through the fully connected layer. Finally, the output of the fully connected layer is processed by softmax function to generate the probability distribution of the prediction. The size of state vector in bi-LSTM is empirically set to 64. Considering that its input contains more information, global

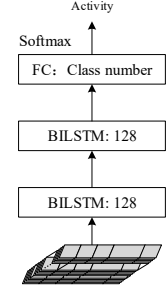

Fig. 4. The architecture of activity classifier.

discriminator uses two bi-LSTM layers, and the size of state vector is set to 128.

### 3.5 Attention Network

Our attention network differentiates the importance of different sensors based on the outputs of feature extractors and the outputs of local discriminators, so that SALIENCE can focus on the sensors with strong feature discrimination and well distribution alignment. Suppose the outputs of feature extractors and local discriminators are $X = \{x_1, x_2, ..., x_n\} \in R^{n \times m}$ and $Y = \{y_1, y_2, ..., y_n\} \in R^{n \times 2}$, respectively, where $n$ is the number of sensors, $m$ is the dimension of feature vectors, $x_i \in R^m$ is the output of feature extractor $i$, and $y_j \in R^2$ is the output of local discriminator $j$. We perform weighted fusion for the outputs of all feature extractors to obtain complete and comprehensive global sensor information $v$ using Equation 5:

$$v = \sum_{i=1}^{n} \alpha_i x_i \qquad (5)$$

where $\alpha_i$ is the attention weight of sensor $i$. Following [50], we obtain the attention weight $\alpha_i$ using Equation 6. The outputs of local discriminators $Y \in R^{n \times 2}$ reflect the alignment degree of feature distributions, which are flatten into $\bar{Y} \in R^{2n}$. Next, we provide query function $Q(\bar{Y})$ and key function $K(X)$ to map $\bar{Y}$ and $X$ to a unified $h$-dimensional latent space (i.e., $Q(\bar{Y}) \in R^h$ and $K(X) \in R^{n \times h}$), respectively. Both functions $Q(.)$ and $K(.)$ are implemented by one layer fully-connected networks. Finally, we use the scaled dot-product of $K(X)$ and $Q(\bar{Y})$ to calculate the attention weights of all sensors:

$$\alpha_1, \alpha_2, ..., \alpha_n = softmax(\frac{Q(\bar{Y})K(X)^T}{\sqrt{h}}) \qquad (6)$$

The attention weight vector is used to differentiate the importance of different sensors, and the weighted feature $v$ is finally input to global discriminator and activity classifier.

### 3.6 Activity Classifier

Activity classifier performs WHAR based on the weighted outputs of all feature extractors. As shown in Fig. 4, two bi-LSTM layers transform the input features and pass the output to a fully connected layer. Finally, softmax function generates the probability distribution of activity prediction based on the output of the fully connected layer. The size of state vector in bi-LSTM is empirically set to 128.

### 3.7 Training

Activity classification loss $\mathcal{L}_C$ and domain discrimination loss $\mathcal{L}_D$ are defined as follows:

$$\mathcal{L}_C = \frac{1}{P}\sum_{i=1}^{P}\left[CE(y_i', y_i^{AC})\right] \quad (7)$$

$$\mathcal{L}_D = \frac{1}{Q}\sum_{i=1}^{Q}\left[\lambda \cdot CE(u_i', u_i^{GD}) + (1-\lambda) \cdot \frac{1}{K}\sum_{j=1}^{K}CE(u_i', u_i^{LD_j})\right] \quad (8)$$

where $P$ is the number of samples from training users, $CE(\cdot)$ represents cross-entropy function, $y_i'$ and $y_i^{AC}$ represent the real label and the prediction label from activity classifier of the $i$-th sample from training users, $Q$ is the total number of samples from training users and the new user, $\lambda$ is a weight factor to balance local and global domain discrimination losses, and $u_i'$ represents the source label of the $i$-th sample, where the label is 0 when the sample comes from training users and 1 when it comes from the new user. In addition, $u_i^{GD}$ and $u_i^{LD_j}$ represent the source prediction of global discriminator and local discriminator $j$, and $K$ is the number of wearable sensors.

Algorithm 1 shows the detailed training process of SALIENCE, which alternately performs the following three steps ($\theta_{FE}$, $\theta_{LD}$, $\theta_{GD}$, $\theta_{AN}$, and $\theta_{AC}$ represent the parameters of feature extractors, local discriminators, global discriminator, attention network, and activity classifier, respectively):

1) Update feature extractors, attention network, and activity classifier to minimize $\mathcal{L}_C$ using the samples from training users (lines 3-5).

2) Update local discriminators and global discriminator to minimize $\mathcal{L}_D$ using the samples from both training users and the new user (lines 6-8). These samples are assigned with source labels.

3) Update feature extractors and attention network to maximize $\mathcal{L}_D$ using the samples from both training users and the new user (lines 9-11). These samples are assigned with source labels.

In steps 2 and 3, an adversarial training strategy is used to align the feature distributions of the data from training users and the new user. Local and global discriminators aim to distinguish whether a sample is from training users or the new user, while feature extractors and attention network try to fool local and global discriminators. By performing steps 2 and 3 alternately, the differences between the feature distributions of the data from training users and the new user are continuously reduced. Feature extractors and attention network eventually learn features that are difficult to distinguish even with well-trained discriminators, so the well-trained WHAR model can be well adapted to the new user. SALIENCE combines local and global discriminators to align the feature distributions. Local discriminators are independent of each other, which provide sensor-level local alignment. Global discriminator aligns the joint data distribution of all sensors to ensure global alignment. Under the guidance of the two objectives, i.e., minimizing activity classification loss $\mathcal{L}_C$ and maximizing domain discrimination loss $\mathcal{L}_D$, attention network focuses the model on sensors with strong feature discrimination and well distribution alignment.

---

**Algorithm 1**: The training process of SALIENCE

**Input:** Dataset of training users $D^{TU}$; Dataset of the new user $D^{NU}$; Learning rate $l$; Number of training iterations $\delta$

**Output:** Trained model

1. Initialize the model with random network parameters
2. **Repeat**
3.    Randomly select a fixed number of samples from $D^{TU}$ to construct mini-batch
4.    Calculate $\mathcal{L}_C$ using Equation 7
5.    $\theta_{FE} \leftarrow \left(\theta_{FE} - l \times \frac{\partial \mathcal{L}_C}{\partial \theta_{FE}}\right)$ ; $\theta_{AN} \leftarrow \left(\theta_{AN} - l \times \frac{\partial \mathcal{L}_C}{\partial \theta_{AN}}\right)$ ; $\theta_{AC} \leftarrow \left(\theta_{AC} - l \times \frac{\partial \mathcal{L}_C}{\partial \theta_{AC}}\right)$
6.    Randomly select a fixed number of samples from $D^{TU}$ and $D^{NU}$ to construct mini-batch
7.    Calculate $\mathcal{L}_D$ using Equation 8
8.    $\theta_{LD} \leftarrow \left(\theta_{LD} - l \times \frac{\partial \mathcal{L}_D}{\partial \theta_{LD}}\right)$; $\theta_{GD} \leftarrow \left(\theta_{GD} - l \times \frac{\partial \mathcal{L}_D}{\partial \theta_{GD}}\right)$
9.    Randomly select a fixed number of samples from $D^{TU}$ and $D^{NU}$ to construct mini-batch
10.   Calculate $\mathcal{L}_D$ using Equation 8
11.   $\theta_{FE} \leftarrow \left(\theta_{FE} + l \times \frac{\partial \mathcal{L}_D}{\partial \theta_{FE}}\right)$; $\theta_{AN} \leftarrow \left(\theta_{AN} + l \times \frac{\partial \mathcal{L}_D}{\partial \theta_{AN}}\right)$
12. **Until** Convergence or reach the number of training iterations $\delta$
13. **Return** Trained model

---

## 4 EXPERIMENTS

### 4.1 Datasets

SALIENCE is evaluated on two public WHAR datasets, i.e., PAMAP2 [38] and OPPORTUNITY [40], both of which provide sufficient multiple sensor data and are widely used for WHAR.

PAMAP2 provides a set of realistic human activity data for WHAR research. Nine users participated in the data collection, including eight males and one female, aged 24 to 32, and they have great differences in height, weight, and heart rate. The participants were asked to wear three inertial measurement units (IMU) and a heart rate monitor, and to perform activities following a protocol containing 12 different daily activities. The sensor signals were collected and labelled with activity labels. A total of nearly 5 hours of labelled data were recorded in PAMAP2, which is publicly available and can be downloaded from [32]. Due to the serious lack of data for one user, the data of remaining eight users are used in the experiments.

OPPORTUNITY provides a set of human activity data in a breakfast scenario. The data were collected in a sensor-rich environment with a realistic way. Four users wearing sensors all over the body participated in the data collection. They were asked to follow a high-level script to perform activities but were free to decide how to achieve these high-level goals. The signals of wearable sensors and environmental sensors were collected simultaneously and labelled with activity labels. Around 6 hours of labelled data were recorded in OPPORTUNITY, which is publicly available and can be downloaded from [30]. In the experiments, only the wearable sensor data are used.

The collected sensor data are usually frame sequences sampled at successive equally spaced points in time, where each frame represents the values acquired by all sensors at a point in time and is labelled with an activity label. These data are first pre-processed with the following steps: 1) Detect and remove invalid values and then complete missing values through linear interpolation; 2) Normalize data by channel to the range of -1 to 1 using Equation 9:

$$x' = 2 \times \frac{x - X_{\min}}{X_{\max} - X_{\min}} - 1 \quad (9)$$

where $x$ is the original value, $X_{\min}$ is the minimum value of the channel where the value is in, $X_{\max}$ is the maximum value of the channel where the value is in, and $x'$ is the normalized value; 3) Segment the continuous frame sequences into a set of equal sized time windows using the sliding window with 1 second overlap. The length of sliding window is set to 2 seconds for the simple activities in PAMAP2, and 10 seconds for the complex activities in OPPORTUNITY referring to [36]; 4) Specify the frame label that appears most in a time window as the label of the time window. The pre-processed time windows are used as the input of the model. The details of the datasets used in the experiments are summarized in Table 1.

### 4.2 Experimental Settings

The source code of SALIENCE is available at GitHub[1]. SALIENCE is implemented based on the PyTorch software library, and the training and testing are conducted under a GPU environment. Adam algorithm [19] is used for model parameter optimization, and the learning rate is set to 0.0001 for OPPORTUNITY and 0.0005 for PAMAP2. The mini-batch size is set to 128. Since the time windows used in the experiments contain relatively many frames, the stride of convolution operation is set to 2.

An experimental protocol is designed to evaluate the performance of a WHAR model on new users. In each evaluation, one user in the dataset serves as the new user and the remaining users serve as training users. The data of training users are used as training set and the data of the new user are randomly split into adaptation set and testing set by a ratio of 0.5:0.5. The labelled data in training set and the unlabelled data in adaptation set are used together for user adaptation. The process is repeated until each new user is evaluated exactly once, and the average result is reported. We set $\lambda = 0.5$ for PAMAP2 and OPPORTUNITY to achieve the balance of local alignment and global alignment.

TABLE 1
The Details of the Datasets Used in the Experiments

|  | PAMAP2 | OPPORTUNITY |
|---|---|---|
| # users | 8 | 4 |
| # activities | 12 | 5 |
| Activities | lying, sitting, standing, walking, running, cycling, Nordic walking, ascending stairs, descending stairs, vacuum cleaning, ironing, rope jumping | relaxing, coffee time, early morning, clean up, sandwich time |
| Sensors | 3 IMUs  1 heart rate monitor | 5 IMUs  2 inertial sensors  12 acceleration sensors |
| Frequency | 100 Hz | 30 Hz |

Accuracy and macro F1-score (abbreviated as F1) are used to measure the performance of a WHAR model. F1 considers both the precision and recall, which is defined formally by Equation 10:

$$F1 = \frac{2}{C} \times \sum_{c=1}^{C} \frac{precision_c \times recall_c}{precision_c + recall_c} \quad (10)$$

where $c$ represents the class index and $C$ represents the number of classes.

### 4.3 Comparison with the State-of-the-Art Models

To demonstrate the competitive performance of SALIENCE, we compare it with the state-of-the-art models, including non-personalized WHAR models, unsupervised user adaptation models for WHAR, and general UDA models, on both datasets. To ensure fairness, all compared models are optimized and follow the same experimental setting as SALIENCE. Note that adaptation set is discarded in the evaluation of non-personalized WHAR models.

Non-personalized WHAR models:

**CNN15** [52]: CNN15 uses a CNN composed of three convolutional layers and two pooling layers to extract feature representations from the input, and then performs activity classification through a softmax classifier.

**DeepConvLSTM** [31]: DeepConvLSTM combines CNN and LSTM for WHAR. The input is first transformed into feature representations by a CNN consisting of four convolutional layers, and then two LSTM layers are used to mine the temporal dependencies in feature representations. Finally, a softmax classifier performs activity classification based on the output of the last LSTM layer.

---
[1] https://github.com/wdkhuans/SALIENCE

**DeepSense** [54]: DeepSense combines CNN and GRU for time series mobile sensing data processing. The input is first split into a series of continuous non-overlapping cells along the time dimension, and the frequency representations of each cell are fed into a CNN to extract feature representations, in which several individual three-layer convolutional subnets separately process the data from different sensors and a global three-layer convolutional subnet fuses the outputs of all individual convolutional subnets. The intra-cell feature representations are then fed into a two-layer GRU, and a softmax classifier performs activity classification based on the output of GRU finally.

Unsupervised user adaptation models for WHAR:

**MMD-Transfer** [16]: MMD-Transfer contains two activity classification networks with shared parameters, one for the samples of training users and the other for the samples of the new user. Each activity classification network exploits two LSTM layers and one fully connected layer to extract feature representations and uses a softmax classifier for activity classification. In addition to activity classification loss, the model introduces MMD to measure the distribution differences between the outputs of LSTM layers in two activity classification networks, and aligns distributions by minimizing MMD.

**HDCNN** [19]: HDCNN contains two symmetric activity classification networks, one for the samples of training users and the other for the samples of the new user. Each activity classification network exploits a CNN composed of two convolutional layers and one fully connected layer to extract feature representations and uses a softmax classifier for activity classification. The model uses KL divergence to measure the distribution differences between the outputs of the corresponding network layers in two activity classification networks, and aligns distributions by minimizing the KL divergence.

**DANN** [7]: DANN consists of a feature extractor, an activity classifier, and a domain discriminator. The activity classifier and the domain discriminator share the output of the feature extractor as their input. The activity classifier and the domain discriminator both contain two fully connected layers and use a softmax classifier for classification, and the feature extractor contains two convolutional layers. The domain discriminator and the feature extractor are adversarially trained to minimize the feature distribution differences between the samples of training users and the samples of the new user.

General UDA models:

**RevGrad** [10]: RevGrad consists of a feature extractor, an activity classifier, and a domain discriminator, which enables the feature extractor to learn domain invariant feature representations in an adversarial way by back-propagating the reverse gradients of the domain discriminator.

**DAN** [25]: DAN exploits a multiple kernel variant of MMD to align feature representations from multiple layers.

**JAN** [26]: JAN learns a transfer model by aligning the joint distributions of the network activation of multiple layers across domains.

**MCD** [44]: MCD aligns the distributions of the source and target domains by exploiting the task-specific decision boundaries.

**CoDATS** [52]: CoDATS aligns the distributions of the source and target domains by exploiting multiple source domains clues.

**AdvSKM** [24]: AdvSKM reforms the MMD metric by considering the non-stationary and non-monotonic statistics in time series distributions.

TABLE 2
The Comparison Between SALIENCE and the State-of-the-Art Models

|  | PAMAP2 | | OPPORTUNITY | |
|---|---|---|---|---|
|  | Accuracy | F1 | Accuracy | F1 |
| CNN15[†] | 0.765* | 0.692* | 0.617* | 0.609* |
| ConvLSTM[†] | 0.802* | 0.733* | 0.656* | 0.662* |
| DeepSense[†] | 0.795* | 0.738* | 0.631* | 0.617* |
| MMD[‡] | 0.799* | 0.741* | 0.668* | 0.658* |
| HDCNN[‡] | 0.830* | 0.744* | 0.717* | 0.711* |
| DANN[‡] | 0.832* | 0.738* | 0.711* | 0.705* |
| RevGrad[‡] | 0.840* | 0.751* | 0.722* | 0.716* |
| DAN[†] | 0.849* | 0.763* | 0.721* | 0.718* |
| JAN[†] | 0.852* | 0.767* | 0.725* | 0.724* |
| MCD[†] | 0.872* | 0.804* | 0.731 | 0.722* |
| CoDATS[†] | 0.828* | 0.757* | 0.728* | 0.731* |
| AdvSKM[†] | 0.810* | 0.714* | 0.729* | 0.721* |
| SALIENCE | **0.894** | **0.828** | **0.744** | **0.741** |

\* means that SALIENCE is statistically superior to the compared model (pairwise t-test at 95% significance level).
† means the baselines have open sourced codes.
‡ means we implemented these baselines by ourselves.

In the experiments, we use the above general UDA models for unsupervised user adaptation in WHAR, and they all keep the same backbone networks as SALIENCE. We follow the published papers to set the parameters of the baselines. For parameters that have not been specified in the published papers, we tune them based on a grid search approach. To statistically measure the significance of the performance differences, pairwise t-tests at 95% significance level are performed between SALIENCE and the state-of-the-art models. The results are shown in Table 2, and the following tendencies can be recognized:

1) SALIENCE outperforms all other models on both datasets. The accuracy and F1 of SALIENCE are 0.021 and 0.024 (on PAMAP2)/0.013 and 0.010 (on OPPORTUNITY) higher than the best values of other models, which shows the competitive performance of our model.

TABLE 3
The Comparison Between SALIENCE and Its Simplified Models

|  | PAMAP2 | | OPPORTUNITY | |
|---|---|---|---|---|
|  | Accuracy | F1 | Accuracy | F1 |
| -base | 0.817* | 0.741* | 0.651* | 0.606* |
| -LD | 0.867* | 0.787* | 0.711* | 0.667* |
| -GD | 0.850* | 0.782* | 0.710* | 0.687* |
| -LD&GD | 0.874* | 0.799* | 0.727* | 0.726* |
| SALIENCE | **0.894** | **0.828** | **0.744** | **0.741** |

\* means that SALIENCE is statistically superior to the compared model (pairwise t-test at 95% significance level).

TABLE 4
The Accuracy of Specific Participants on PAMAP2

| Test User | -base | -LD | -GD | -LD&GD | SALIENCE |
|---|---|---|---|---|---|
| User1 | 0.715 | 0.787 | 0.762 | 0.800 | 0.776 |
| User2 | 0.802 | 0.886 | 0.881 | 0.907 | 0.927 |
| User3 | 0.900 | 0.87 | 0.883 | 0.865 | 0.868 |
| User4 | 0.906 | 0.902 | 0.923 | 0.921 | 0.947 |
| User5 | 0.847 | 0.875 | 0.883 | 0.892 | 0.900 |
| User6 | 0.887 | 0.871 | 0.919 | 0.903 | 0.930 |
| User7 | 0.927 | 0.912 | 0.939 | 0.929 | 0.911 |
| User8 | 0.550 | 0.837 | 0.612 | 0.770 | 0.892 |
| **Overall** | **0.817** | **0.867** | **0.850** | **0.874** | **0.894** |

2) Non-personalized WHAR models perform worse than personalized WHAR models, which demonstrates the effectiveness of unsupervised user adaptation for WHAR. By using easily acquired unlabelled data of the new user for unsupervised user adaptation, the performance of WHAR models on the new user can be significantly improved.

3) There is no significant performance gap between general UDA based personalized WHAR models and existing unsupervised user adaptation based personalized WHAR models. This is mainly because that most existing unsupervised user adaptation methods for WHAR do not consider the characteristics of WHAR task, but directly use general UDA methods. The design of our model considers the characteristics of multiple sensor data to further improve the performance.

4) The comparison between SALIENCE and other personalized WHAR models demonstrates the superiority of our model. The results suggest that the proposed alignment scheme and attention mechanism can help our model better handle multiple sensor data, thereby improving the performance.

## 4.4 Comparison with Simplified Models

To investigate the effectiveness of the proposed alignment scheme and attention mechanism for multiple wearable sensors based WHAR, the comparison experiments between SALIENCE and its simplified models are conducted on both datasets, where these simplified models are obtained by removing some key components from SALIENCE. To ensure fairness, all simplified models are optimized and follow the same experimental setting as SALIENCE.

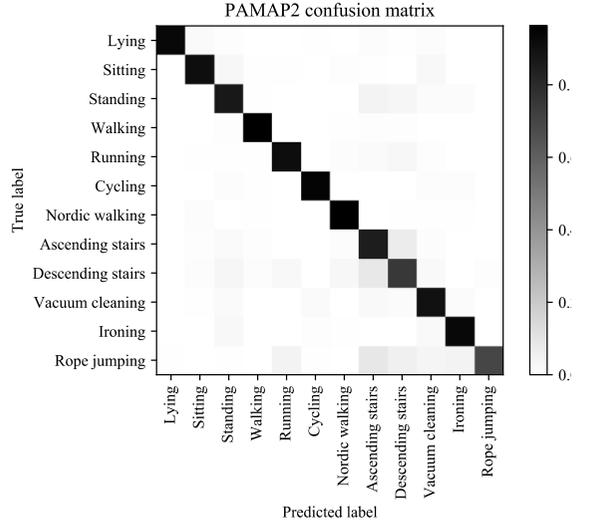

Fig. 5. The confusion matrix of SALIENCE on PAMAP2.

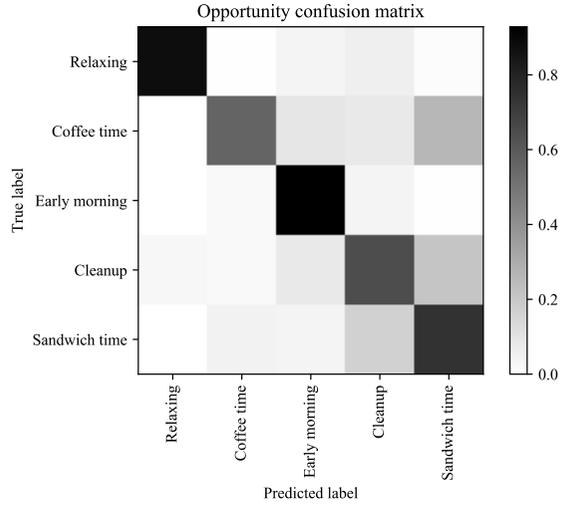

Fig. 6. The confusion matrix of SALIENCE on OPPORTUNITY.

Simplified models:

**SALIENCE-base**: SALIENCE-base is a baseline model of SALIENCE without user adaptation. It consists of feature extractors and activity classifier with the same structure as in SALIENCE. Note that adaptation set is discarded in its evaluation.

**SALIENCE-LD**: SALIENCE-LD removes global discriminator and attention network from SALIENCE, and the rest is consistent with it.

**SALIENCE-GD**: SALIENCE-GD removes local discriminators and attention network from SALIENCE, and the rest is consistent with it.

**SALIENCE-LD&GD**: SALIENCE-LD&GD removes attention network from SALIENCE, and the rest is consistent with it.

To statistically measure the significance of the performance differences, pairwise t-tests at 95% significance level are performed between SALIENCE and its simplified models. The results are shown in Table 3, which show the following tendencies:

1) The performance of SALIENCE-base is obviously worse than that of other models with unsupervised user adaptation, which demonstrates the effectiveness of unsupervised user adaptation for WHAR. Unsupervised user adaptation aligns the feature distributions of the data from training users and the new user, so a well-trained WHAR model can be well adapted to the new user.

2) SALIENCE-LD&GD outperforms SALIENCE-LD and SALIENCE-GD, which shows that combining local and global alignments can achieve better user adaptation for multiple wearable sensors based WHAR.

3) SALIENCE is superior to SALIENCE-LD&GD, which suggests that the proposed attention mechanism can focus the model on the sensors with strong feature discrimination and well distribution alignment, thereby improving model performance. The comparison between SALIENCE and its simplified models implies that combining the proposed alignment scheme and attention mechanism can achieve better performance.

Next, we investigate the detailed performance of the above models. Specifically, we investigate the accuracy of specific users on PAMAP2, and the results are shown in Table 4. It can be seen that on User 8, SALIENCE-LD&GD and SALIENCE have significant accuracy improvement over SALIENCE-GD and SALIENCE-LD&GD, respectively. User 8 is the only user that uses left hand as dominant hand, and the data distribution shift is much higher. The results demonstrate that the larger the discrepancy between training users and the new user is, the greater the benefits of introducing local discriminators and the proposed attention mechanism are.

## 4.5 Model Investigation

We give the confusion matrices of activity recognition on PAMAP2 and OPPORTUNITY in Fig. 5 and Fig. 6, respectively. In PAMAP2, "ascending stairs" and "descending stairs" tend to be misclassified to each other, because the two activities have similar body actions. In OPPORTUNITY, "Coffee time", "Sandwich time", and "Clean" tend to be misclassified with each other. Here, the difference between "Coffee time" and "Sandwich time" is vague, and the three activities are all characterized by arm and hand actions.

Fig. 7 presents the feature distributions of the new user (User 1) and training users (the other users). The dataset is PAMAP2, features are the output of activity classifier before the softmax layer, and t-SNE is used to visualize the features. It can be found that SALIENCE achieves much smaller feature distribution discrepancy than SALIENCE-base.

## 4.6 Case Study

In this section, we conduct a case study to investigate the distribution of attention weights of different sensors. Fig. 8 (a) and (b) shows the average attention weights of different sensors in the recognition of activity "ironing" for the new user 1 (User 6) and the new user 2 (User 7) in PAMAP2, respectively, and the corresponding average output differences (i.e., the difference between the two output values of the softmax function of activity classifier) of local discriminators, where a larger output difference means a worse alignment.

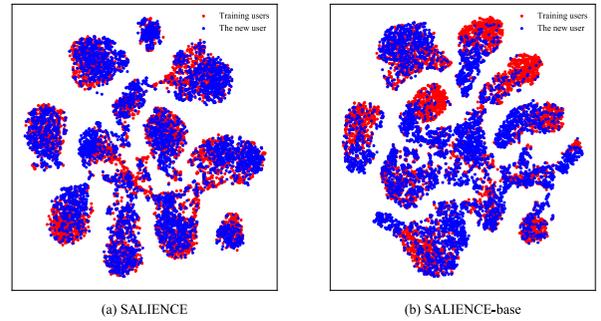

Fig. 7. Visualization of the feature distributions of the new user and training users using t-SNE.

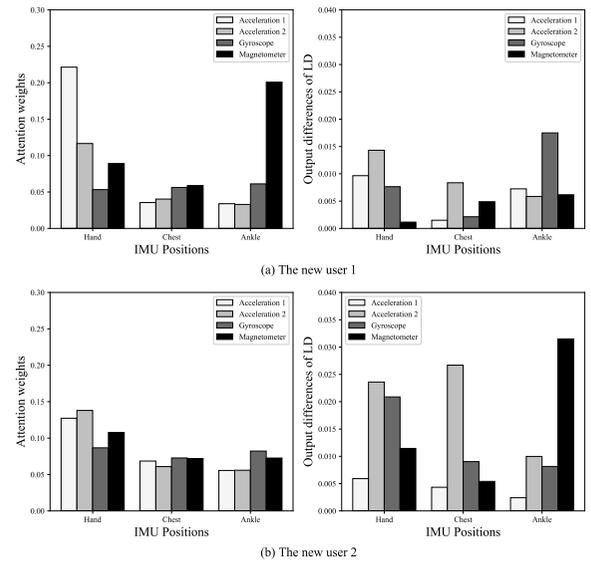

Fig. 8. A case study of attention weights and output differences of local discriminators.

We found that in general, the attention weight distributions of the new user 1 and the new user 2 are relatively consistent, where weights are heavily distributed on the IMU on the hand. It is intuitive since "ironing" is an activity mostly using arms and hands. Meanwhile, we found that there are differences in the attention weights of some sensors. For example, for Magnetometer (ankle), the attention weight of the new user 1 is much higher than that of the new user 2. It is because for the new user 2, the alignment of Magnetometer (ankle) is poor, as shown in Fig. 8 (b), causing attention network to pay less attention to the sensor. In this way, our attention network helps the activity classifier of SALIENCE focus on the sensors with strong feature discrimination and well distribution alignment.

## 5 Conclusions and Future Work

The use of multiple wearable sensors in WHAR supports recognizing more diverse and complex activities and brings more accurate recognition results, but also poses a huge challenge for unsupervised user adaptation. In this paper, we propose SALIENCE model to address different transferbilities of different sensors, which aligns the data of each sensor separately to achieve local alignment, while uniformly aligning the data of all sensors to ensure global alignment. Furthermore, an attention mechanism is proposed to focus our model on the sensors with strong feature discrimination and well distribution alignment to further improve the performance and generalization. Experimental results demonstrate the effectiveness of SALIENCE on two public WHAR datasets.

In the future, we will extend our work in the following aspects. First, we will extend SALIENCE to support both unsupervised and supervised user adaptation. Second, we will consider the conditional distribution alignment along with marginal distribution alignment to achieve better UDA. Third, we will evaluate our model on more datasets to make the results more convincing and explore the impact of key hyperparameters.

## Acknowledgment

This work is supported by the National Key Research and Development Program of China (No. 2018YFB0505000), the Joint Funds of the National Natural Science Foundation of China (No. U1936215), the Key R&D Plan of Zhejiang Province (No. 2021C01117), and the Industrial Internet Innovation Development Project (No. TC200H01V).

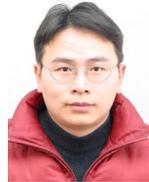
**Ling Chen** received the B.S. and Ph.D. degrees in computer science from Zhejiang University, China, in 1999 and 2004, respectively. He is currently an associate professor with the College of Computer Science and Technology, Zhejiang University, China. His research interests include ubiquitous computing, AI, pattern recognition, and human computer interaction.

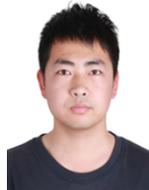
**Yi Zhang** received his B.S. degree in computer science and technology from Northwestern Polytechnical University, China, in 2017, M.S. degree in computer science and technology from Zhejiang University, China, in 2020. His research interests include wearable computing and human activity recognition.

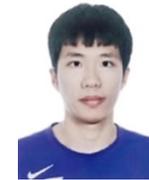
**Shenghuan Miao** received his B.S. degree in information engineering from Zhejiang University of Technology, China, in 2021. He is currently a Ph.D. candidate in the College of Computer Science and Technology, Zhejiang University, China. His research interests include wearable computing and human activity recognition.

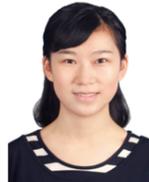
**Sirou Zhu** received her B.S. degree in software engineering from Zhejiang University, China, in 2019. She is currently a M.S. candidate in the School of Computer Science, Carnegie Mellon University, USA. Her research interests include optimization and human computer interaction.

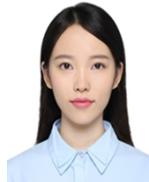
**Rong Hu** received her B.S. degree in applied mathematics from Zhejiang University of Technology, China, in 2019. She is currently a Ph.D. candidate in the College of Computer Science and Technology, Zhejiang University, China. Her research interests include wearable computing and human activity recognition.

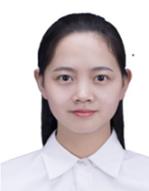
**Liangying Peng** received her B.S. degree in software engineering from Nankai University, China, in 2014, Ph.D. degree in computer science and technology from Zhejiang University, China, in 2020. Her research interests include wearable computing and human activity recognition.

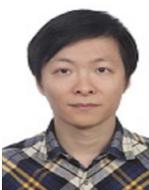
**Mingqi Lv** received his Ph.D. degree in computer science from Zhejiang University, China, in 2012. He is currently an associated professor with the College of Computer Science and Technology, Zhejiang University of Technology, China. His research interests include spatiotemporal data mining and ubiquitous computing.